\documentclass[journal]{IEEEtran}

\usepackage{amssymb}
\usepackage[utf8]{inputenc}
\usepackage[T1]{fontenc}
\usepackage{array,multirow}
\usepackage{rotating}
\usepackage{color}
\usepackage{arydshln}
\usepackage[numbers,sort&compress]{natbib}
\usepackage{url}

\newcolumntype{P}[1]{>{\centering\arraybackslash}p{#1}}

\usepackage[]{units}

\IEEEoverridecommandlockouts                              

\begin{document}
    
\title{Automatic sleep monitoring using ear-EEG}
\author{Takashi~Nakamura${\IEEEauthorrefmark{1}}$,
	 Valentin~Goverdovsky${\IEEEauthorrefmark{1}}$,~\IEEEmembership{Member,~IEEE,}
	 Mary~J.~Morrell${\IEEEauthorrefmark{2}}$,
	 and~Danilo~P.~Mandic${\IEEEauthorrefmark{1}}$,~\IEEEmembership{Fellow,~IEEE}
\thanks{${\IEEEauthorrefmark{1}}$TN, VG and DPM are with Department of Electrical and Electronic Engineering, Imperial College London,
London, SW7 2AZ, United Kingdom, \{takashi.nakamura14, goverdovsky, d.mandic\}@imperial.ac.uk}%
        \thanks{${\IEEEauthorrefmark{2}}$MJM is with Sleep and Ventilation Unit, National Heart and Lung Institute, Imperial College London, and NIHR Respiratory Disease Biomedical Research Unit at the Royal Brompton and Harefield NHS Foundation Trust, and Imperial College London London, SW3 6NP, United Kingdom, m.morrell@imperial.ac.uk}}


\twocolumn[
\begin{@twocolumnfalse}
\maketitle
\noindent \normalsize{\textbf{Abstract} The monitoring of sleep patterns without patient's inconvenience  or  involvement of a medical specialist is a clinical question of significant importance. To this end, we propose an automatic sleep stage monitoring system based on an affordable, unobtrusive, discreet, and long-term wearable in-ear sensor for recording the Electroencephalogram (ear-EEG).  The selected features for sleep pattern classification from a single ear-EEG channel include the spectral edge frequency (SEF) and multi-scale fuzzy entropy (MSFE), a structural complexity feature. In this preliminary study, the manually scored hypnograms from simultaneous scalp-EEG and ear-EEG recordings of four subjects are used as labels for two analysis scenarios: 1) classification of ear-EEG hypnogram labels from ear-EEG recordings and 2) prediction of scalp-EEG hypnogram labels from ear-EEG recordings. We consider both 2-class and 4-class sleep scoring, with the achieved accuracies ranging from \unit[78.5]{\%} to \unit[95.2]{\%} for ear-EEG labels predicted from ear-EEG, and \unit[76.8]{\%} to \unit[91.8]{\%}  for scalp-EEG labels predicted from ear-EEG. The corresponding kappa coefficients,  which range from 0.64 to 0.83 for Scenario 1 and from 0.65 to 0.80 for Scenario 2, indicate  a Substantial to Almost Perfect agreement, thus proving the feasibility of in-ear sensing for sleep monitoring in the community.}

\vspace{0.5cm}
\vspace{0.5cm}
\end{@twocolumnfalse}
]
{
  \renewcommand{\thefootnote}%
    {\fnsymbol{footnote}}
  \footnotetext[1]{Takashi Nakamura, Valentin Goverdovsky, and Danilo P. Mandic are with Department of Electrical and Electronic Engineering, Imperial College London,
London, SW7 2AZ, United Kingdom, \{takashi.nakamura14, goverdovsky, d.mandic\}@imperial.ac.uk}
 \footnotetext[2] {Mary J. Morrell is with Sleep and Ventilation Unit, National Heart and Lung Institute, Imperial College London, and NIHR Respiratory Disease Biomedical Research Unit at the Royal Brompton and Harefield NHS Foundation Trust, and Imperial College London, London, SW3 6NP, United Kingdom, m.morrell@imperial.ac.uk}
}
                                                          
    \thispagestyle{empty}
    \pagestyle{empty}
\section{Introduction}
Sleep is an essential process in the internal control of the state of body and mind and its quality is strongly linked with a number of cognitive and health issues, such as stress, depression and memory \cite{Maquet2001}. 
For clinical diagnostic purposes, polysomnography (PSG) has been extensively utilised which is based on a multitude of physiological responses including the electroencephalogram (EEG), electrooculogram (EOG), and electromyogram (EMG).
While the PSG is able to faithfully reflect human sleep patterns, both the recording and scoring process are expensive as this involves an overnight stay in a specialised clinic and time-consuming manual scoring by a medically trained person. 
In addition, hospitals are unfamiliar environments for patients, which compromises the reliability of the observed sleep patterns. 
In other words, the conventional recording process is not user-centred and not ideal for long-term sleep monitoring. 

With the advance in wearable physiological monitoring devices, it has become possible to monitor some of sleep-related physiological responses out of the clinic.
The next step towards sleep care in the community is therefore to monitor sleep-related physiological signals in an affordable way, at home, and over long periods of time, together with automatic detection of sleep patterns (sleep scoring) without the need for a trained medical expert.
Indeed, consumer technologies are becoming increasingly popular for the self-monitoring of sleep \cite{Ko2015}, and include both mobile apps and wearable devices. While such technologies aim to assess `sleep quality' and are affordable, these are typically not direct measures of neural activity, and instead measure indirect surrogates of sleep such as limb movement \cite{Rosenzweig2014}. 

Another fast developing aspect of sleep research is on automatic sleep scoring, with the aim to replace the time-consuming manual scoring of sleep patterns from full PSG with computer software. 
The manual sleep scoring is performed through a visual interpretation of 30-second PSG recordings, and based on well-established protocols such as the manual of the American Academy of Sleep Medicine (AASM) \cite{iber2007aasm}. The diagnostically relevant sleep stages include: wake (W), non-rapid eye movement (NREM) Sleep Stage 1 (N1), NREM Stage 2 (N2), NREM Stage 3 (N3), and REM \cite{Carskadon2005}. 
Automatic sleep stage scoring employs machine learning and pattern recognition algorithms, and it is now possible to achieve up to \unit[90]{\%} accuracy of classification between the W, N1, N2, N3 and REM sleep stages from a single channel EEG \cite{Koley2012,DaSilveira2016}. Publicly available resources to evaluate automatic sleep stage classification algorithms include the Sleep EDF database \cite{PhysioNet}. A single channel EEG montage is therefore a prerequisite for a medical-grade wearable system and for benchmarking new developments against existing solutions.

\begin{center}
    \begin{figure}[htbp]
    \centering
		\includegraphics[clip, width=\linewidth]{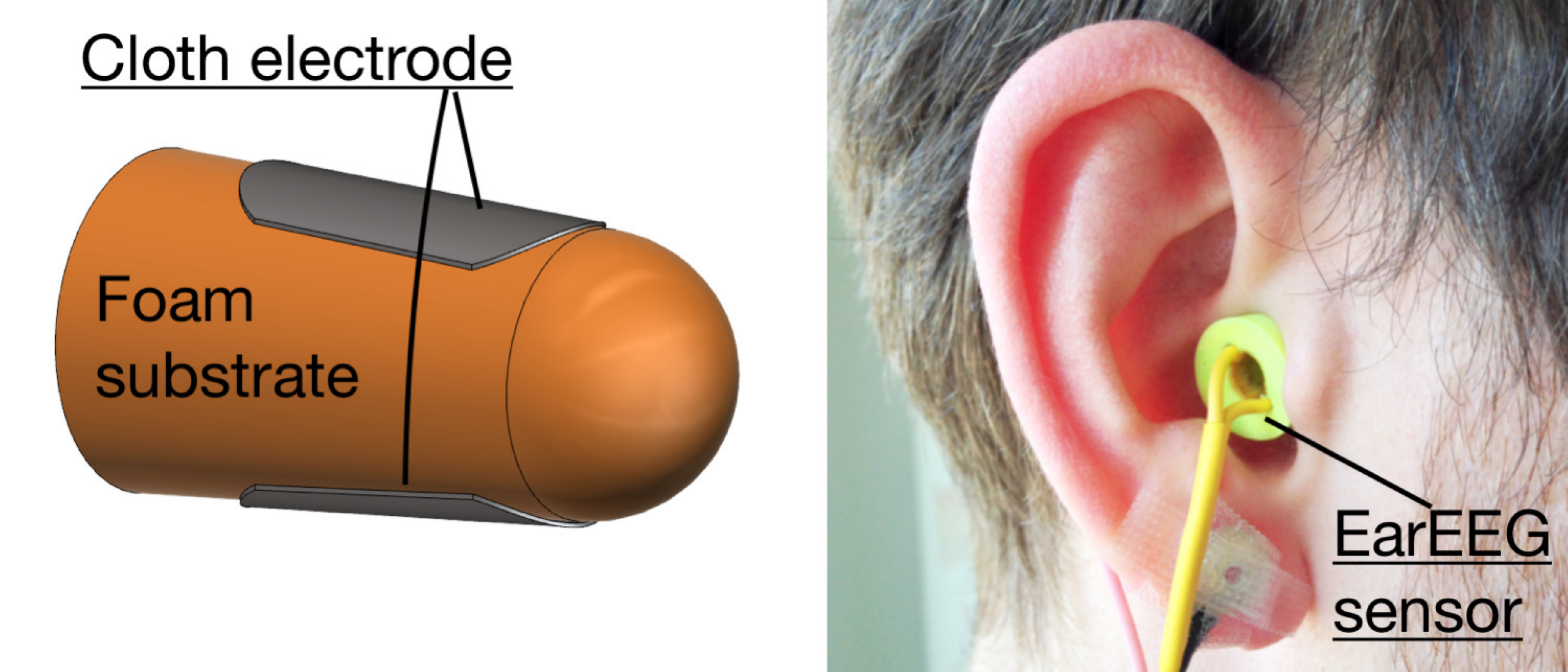}
   	\caption{The in-ear sensor used in our study. Left: Wearable in-ear sensor with two flexible electrodes. Right: Placement of the generic earpiece.}
	\label{fig:earpiece_figure}
	\end{figure}
	\end{center}

More recent approaches for sleep monitoring aim to move beyond actigraphy and develop advanced multimodal sensors and wearable devices. In this direction, Le {\it et al.} introduced a wireless wearable sensor to monitor vectorcardiography (VCG), ECG, and respiration for detecting obstructive sleep apnea in real time \cite{Le2013}. Using a wearable in-ear EEG sensor (ear-EEG) \cite{Looney2012}, Looney {\it et al.} monitored fatigue, while our recent work evaluated sleep stages during nap episodes from a viscoelastic in-ear EEG sensor \cite{Looney2016a}, see Figure \ref{fig:earpiece_figure}. Stochholm {\it et al.} monitored EEG from inside ear canal during sleep, and undertook automatic sleep stage classification \cite{Stochholm2016}.

The in-ear sensing technology has been proven to provide sufficiently good EEG signal for brain-computer interface applications with steady-state responses \cite{Looney2012, Kidmose2013a,Yu-TeWang2015}, and has more recently been used for monitoring other physiological responses, such as cardiac activity \cite{Goverdovsky2015, goverdovsky2016hearables}. Such a wearable system is designed to be comfortable over long periods of time and with the electrodes are firmly placed inside of ear canal, which ensures good quality of recordings. Even though amplitude of ear-EEG is smaller than that of scalp-EEG, the signal-to-noise ratio (SNR) was found to be similar \cite{Looney2012, Kidmose2013a, Goverdovsky2016}. In a sleep monitoring scenario, in-ear wearable sensors have the following advantages:
\begin{itemize}
	\item {\it Affordability and unobtrusiveness}: Our latest sensor (generic earpiece) is made from viscoelastic material \cite{Goverdovsky2016}, such as those used in standard earplugs, see Figure \ref{fig:earpiece_figure}.
	\item {\it User-centred nature}: Users are able to insert the sensor by themselves as when wearing earplugs. The device is comfortable to wear and does not disturb sleep. 
	\item  {\it Robustness}: The sensor expands after the insertion and  maintains a stable interface with the ear canal, and is thus not likely to dislodge during sleep.
\end{itemize}
\begin{center}
    \begin{figure}[htbp]
    		\includegraphics[clip, width=\linewidth]{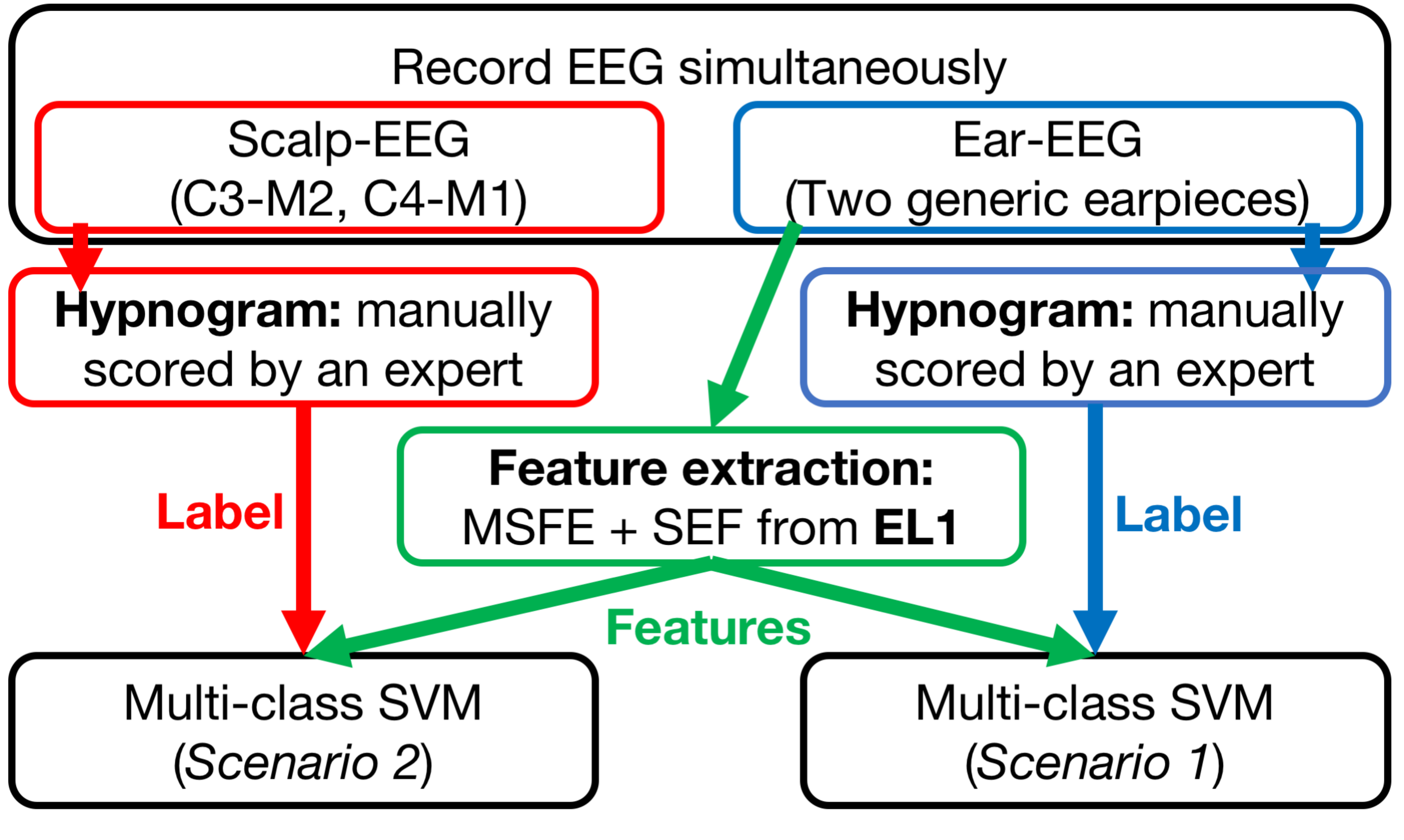}
    	\caption{Analysis framework for the feasibility of ear-EEG in sleep research.}
	\label{fig:Problem_formulation}
	\end{figure}
	\end{center}

In order to examine the feasibility of sleep monitoring with the ear-EEG sensor, we set out to establish a comprehensive cross-validation between standard clinical scalp-EEG recording and our own ear-EEG recordings. As a step in this direction, in our recent study \cite{Looney2016a}, we performed simultaneous sleep monitoring using both scalp- and ear-EEG data channels, and reported substantial agreement between the corresponding hypnograms, manually scored by a trained clinical person. Here, we embark upon the proof-of-concept results in \cite{Looney2016a} and make another step towards fully wearable sleep monitoring through automatic sleep stage classification. The sleep-related EEG-patterns were obtained from both scalp and inside the ear simultaneously, using a stationary data acquisition unit. For rigour, the scalp- and ear-EEG automatic scoring procedures were validated for the following scenarios:

\begin{enumerate}
	\item Agreement between the hypnogram scored \emph{manually based on ear-EEG patterns} and the automatically predicted label based on ear-EEG patterns. (\emph{Scenario 1}).
	\item Agreement between the hypnogram scored \emph{manually based on scalp-EEG patterns} and the automatically predicted label based on ear-EEG patterns. (\emph{Scenario 2}).
	\end{enumerate}

Figure \ref{fig:Problem_formulation} illustrates the proposed analysis framework. The results are benchmarked against the results in \cite{Looney2016a} where both the scalp- and ear-EEG hypnograms were scored manually. In this way, we establish a proof-of-concept for the feasibility of ear-EEG in automatic scoring of sleep patterns out-of-clinic and in the community. 
	
%
%
\section{Methods}

\subsection{Data Acquisition}

The EEG recordings were conducted at Imperial College London between May 2014 and March 2015 under the ethics approval, ICREC 12\_1\_1, Joint Research Office at Imperial College London. Four healthy male subjects (age: 25 - 36 years) without history of sleep disorders participated in the recordings. All participants were instructed to reduce their sleep to less than 5 hours the night before, and agreed to refrain from consuming caffeine and napping on the recording day.
The four scalp-EEG channels C3, C4, A1 and A2 (according to international 10-20 system), were recorded using standard gold-cup electrodes. The forehead was used for the ground, and the standard configurations for sleep scoring were utilised (i.e. C3-A2 and C4-A1). The ear-EEG was recorded from both the left and right ear, and the ear-EEG sensor was made based on a viscoelastic earplug with two cloth electrodes \cite{Goverdovsky2016}, as shown in Figure \ref{fig:earpiece_figure}. Earwax was removed from the ear canals, and the sensor expanded after the insertion, to conform to the shape of the ear canal. The reference gold-cup standard electrodes were attached behind the ipsilateral mastoid and the ground electrodes were placed on the ipsilateral helix. Both scalp-EEG and ear-EEG were recorded simultaneously using the g.tec g.USBamp amplifier with 24-bit resolution, at a sampling frequency were $fs = $ \unit[1200]{Hz}. 

The participants seated in a comfortable chair in a dark and quiet room. The duration of recording was 45 minutes, while to increase the number of transitions between the wake and sleep stage, a loudspeaker played \unit[10]{s} abrupt noise at random intervals. 

\begin{center}
    \begin{figure}[htbp]
    		\includegraphics[clip, width=\linewidth]{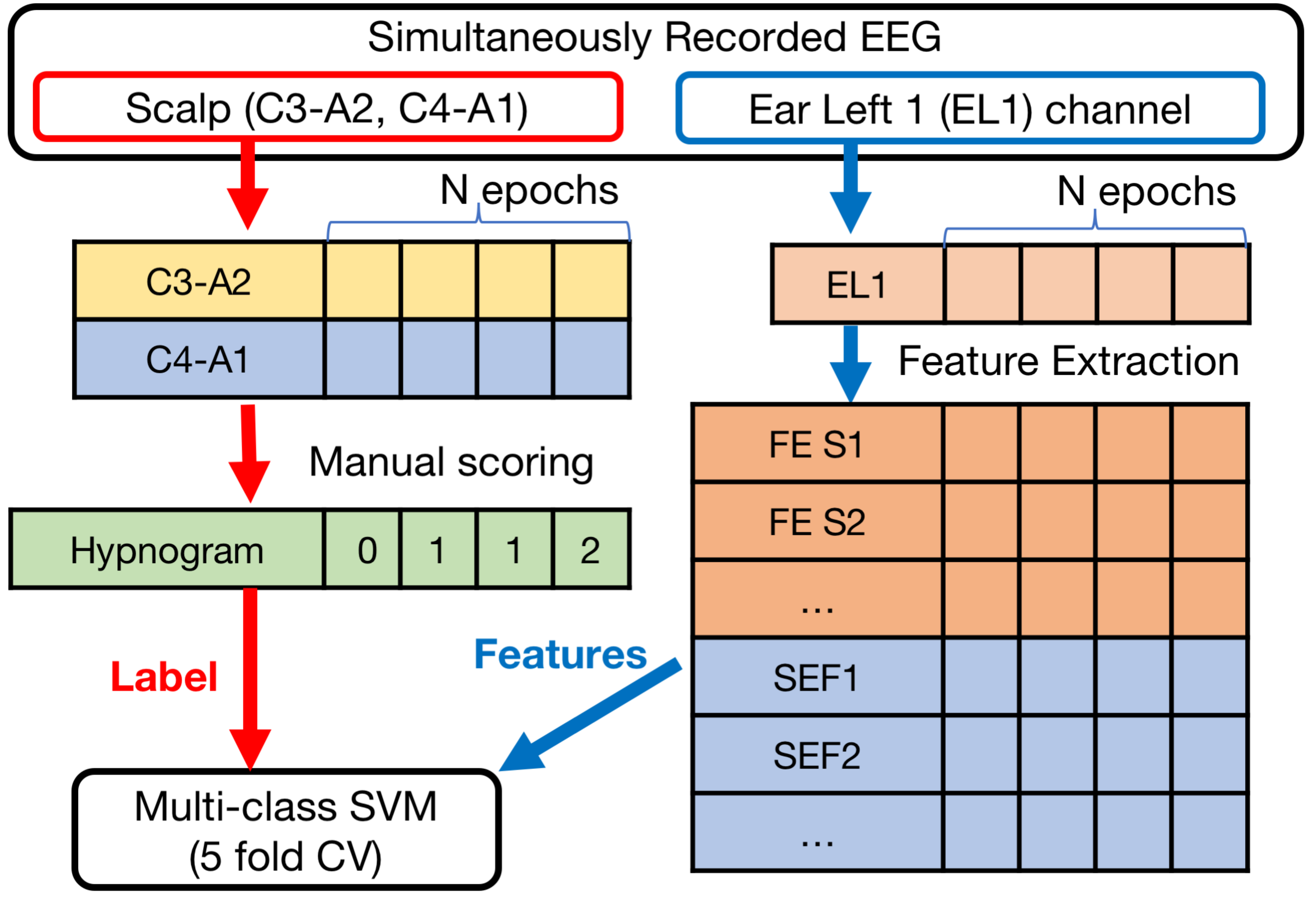}
    	\caption{Flowchart for a sleep stage prediction framework adopted in this study (\emph{Scenario 2}).}
	\label{fig:HypRes}
	\end{figure}
\end{center}
\subsection{Sleep stage scoring}

For scalp-EEG, a 4th-order Butterworth bandpass filter with passband [$1 - 20$] Hz was applied to two bipolar EEG configurations (i.e. C3-A2 and C4-A1). Due to low-frequency interference in ear-EEG channels, the low cutoff frequency was set to \unit[1]{Hz} for the Subject 1 and 3, and \unit[2]{Hz} for the Subject 2 and 4. Next, the ear-EEG amplitudes were normalised to the same range as those of scalp-EEG, and both scalp-EEG and ear-EEG were manually scored by a clinical expert, who had six years of experience in EEG-based sleep stage scoring. The processed EEG data was blinded and the epoch-based manual sleep scoring was performed according to the American Academy of Sleep Medicine (AASM) criteria \cite{iber2007aasm}. The epoch size was set to \unit[30]{s}, therefore 90 epochs were scored in each recording. 


\subsection{Pre-processing for automatic stage classification}

For automatic sleep stage classification, we considered the recorded EEG from the left ear channel 1 (EL1), for a fair comparison with automatic scoring algorithms for a single EEG channel montage in the literature. First, the data was downsampled to \unit[200]{Hz}, and the epochs with the amplitudes of more than \unit[$\pm$400]{$\mu$V} were removed from subsequence analyses. The data were then bandpass filtered with the passband of  [$ 0.5 - 30$] Hz. The pre-processing resulted in a loss of approximately \unit[20]{\%} of the data, and eventually 293 (hypnogram-based on scalp-EEG, W:67, N1:46, N2:140, N3:40, and hypnogram-based on ear-EEG, W:52, N1:49, N2:162, N3:30) epochs were used for the classification. 

\subsection{Feature Extraction} 

After the pre-processing, two types of features were extracted from each epoch of the EEG. These were the same as those in the latest automatic sleep stage classification results based on the Sleep EDF database \cite{Nakamura2017}, and included: 1) a frequency domain feature - spectral edge frequency (SEF), and 2) a structural complexity feature - multi-scale entropy (MSE).

\subsubsection{Frequency domain features}
\begin{center}
    \begin{figure}[htbp]
		\includegraphics[clip, width=\linewidth]{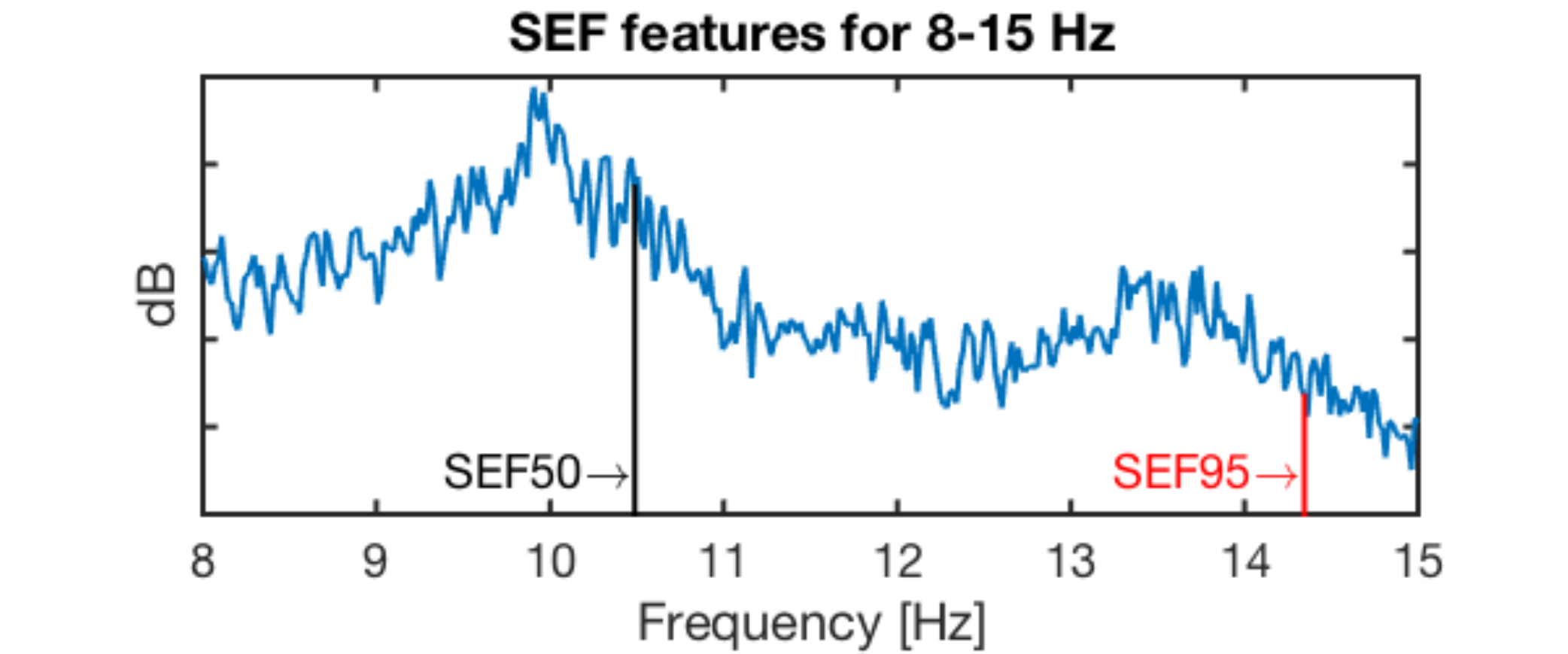}
    	\caption{Spectral edge frequency (SEF) features for \unit[8 - 15]{Hz}. The symbol SEF$50$ denotes the lowest frequency below which \unit[50]{\%} of the total power in a considered frequency band is contained ({\it cf.} SEF$95$ for \unit[95]{\%} of total power).}
	\label{fig:SEF_model}
	\end{figure}
\end{center}

\begin{center}
    \begin{figure}[htbp]
		\includegraphics[clip, width=\linewidth]{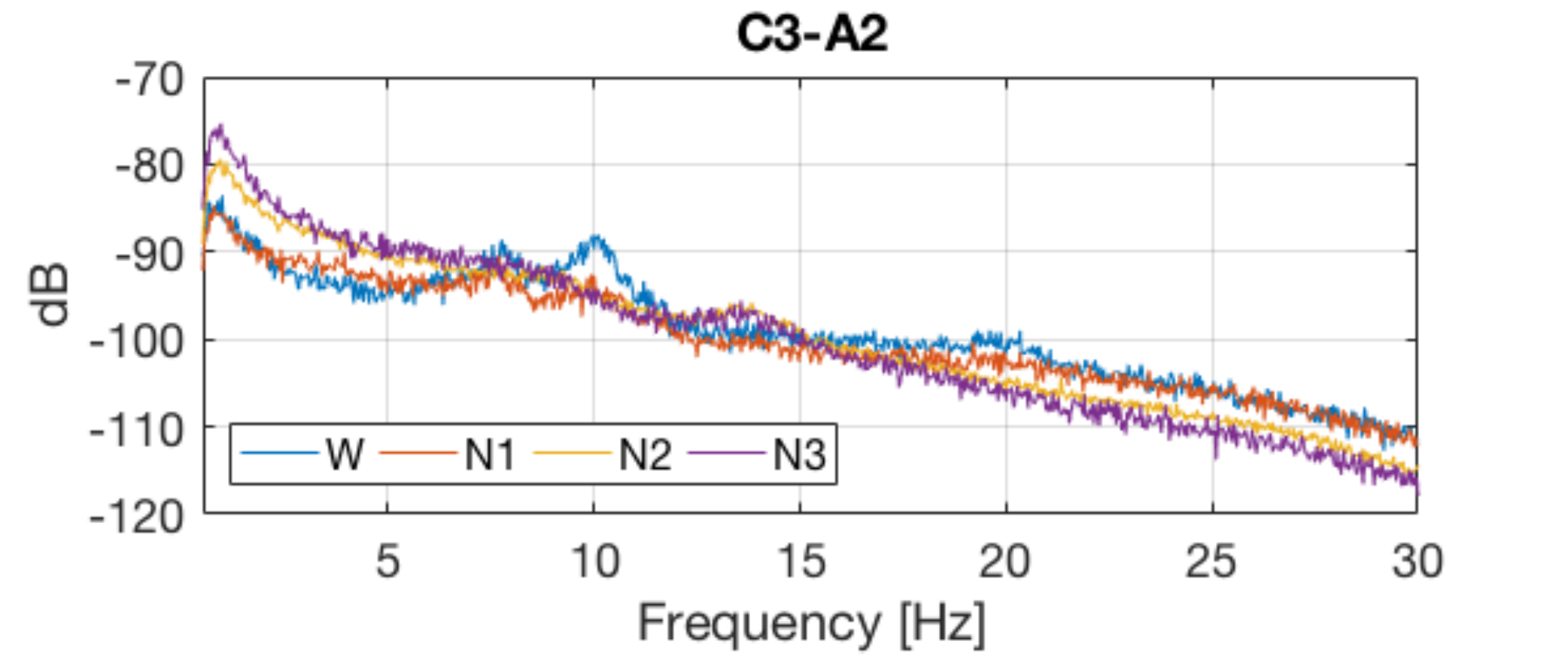}
    		\includegraphics[clip, width=\linewidth]{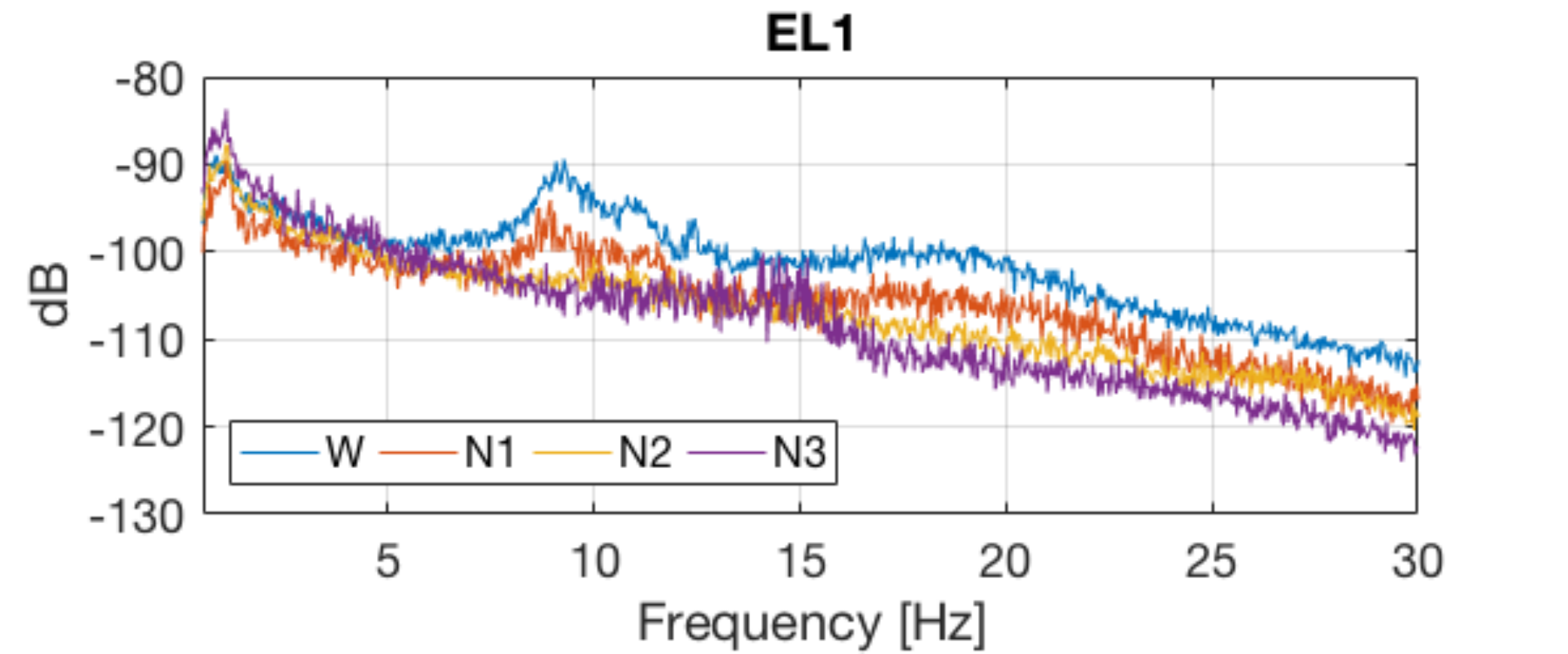}
    	\caption{Power spectral density for the scalp C3-A2 montage (top) and for the in-ear EEG channel EL1 (bottom).}
	\label{fig:FFT}
	\end{figure}
\end{center}

\begin{center}
    \begin{figure}[htbp]
		\includegraphics[clip, width=\linewidth]{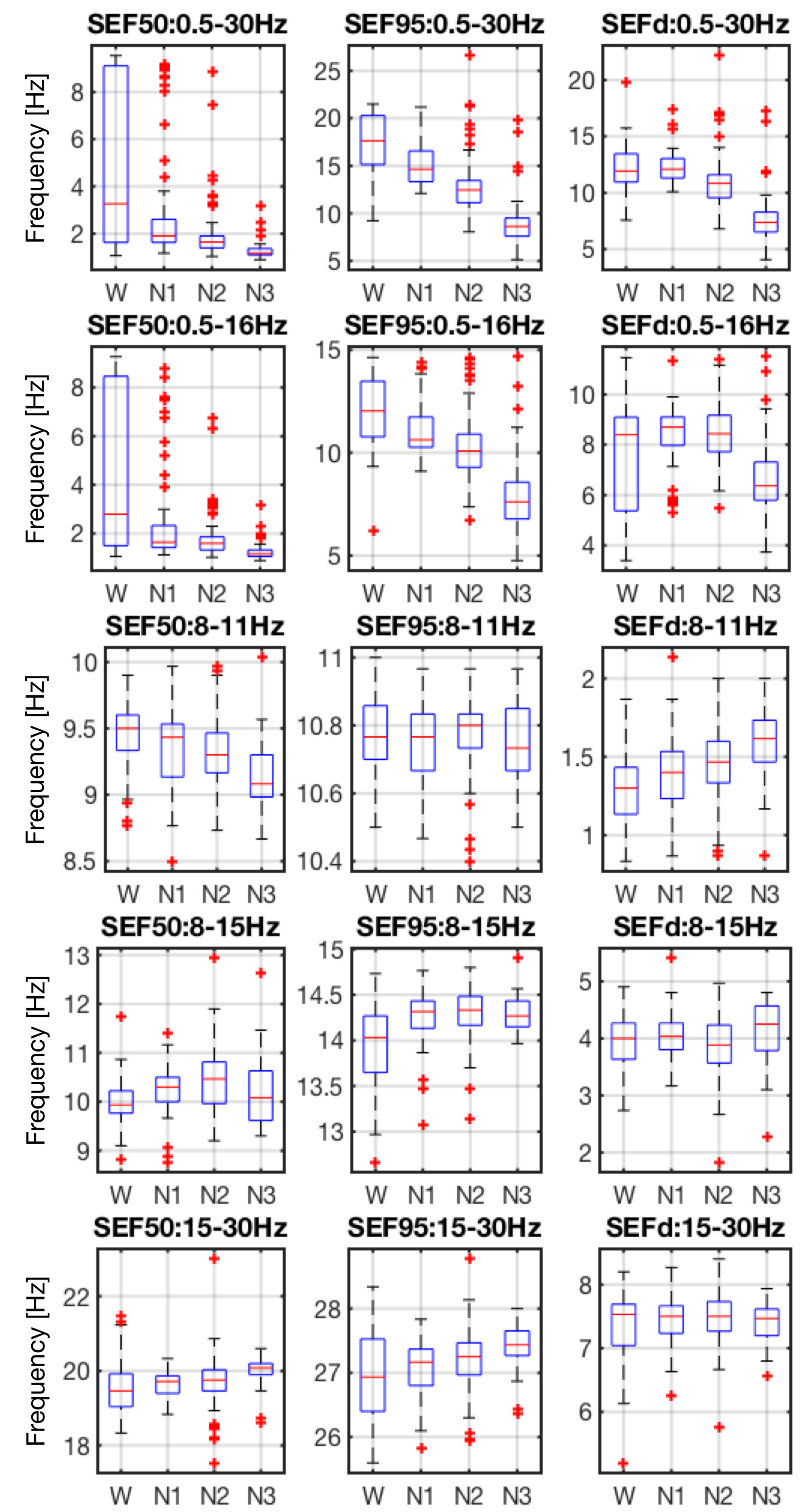}
    	\caption{The frequency domain SEF$50$, SEF$95$, and SEF$d$ features of the $\delta - \beta$, $\delta - \alpha$, $\alpha_l$, $\alpha$, and $\beta$ band power from the in-ear EEG channel EL1. The features were averaged over all epochs and subjects.}
	\label{fig:SEF}
	\end{figure}
\end{center}

The \unit[$r$]{\%} of spectral edge frequency (SEF$r$) is calculated as the $r$th percentile of the total power calculated from power spectral density, as illustrated in Figure \ref{fig:SEF_model}. Figure \ref{fig:FFT} illustrates the power spectral density for the scalp C3-A2 (top) and in-ear EL1 (bottom) channels for different sleep stages, labeled manually based on scalp-EEG patterns. Observe that the spectral patterns \cite{Silber2007} in scalp-EEG and ear-EEG are similar: the alpha (\unit[8 - 13]{Hz}) band power in the Wake condition, a slightly smaller alpha power in N1 sleep, and the stronger power of the delta ($<$ \unit[2]{Hz}) band towards deep sleep. We next obtained the SEF$50$ and SEF$95$ features for the following frequency bands: $\delta - \beta$ = \unit[0.5 - 30]{Hz}, $\delta - \alpha$ = \unit[0.5 - 16]{Hz}, $\alpha_l$ = \unit[8 - 11]{Hz}, $\alpha$ = \unit[8 - 15]{Hz}, and $\beta$ = \unit[16 - 30]{Hz}. In addition, the SEF$d$ feature was calculated as the difference between SEF$95$ and SEF$50$, that is SEF$d$ = SEF$95$ - SEF$50$; so that 15 SEF features were obtained from the in-ear EL1 channel. Figure \ref{fig:SEF} shows the boxplots of SEF features in different frequency bands for the EL1 channel and for each sleep stage, averaged over all epochs and subjects. Observe the consistent spread of SEF features.

\subsubsection{Structural complexity features}
\begin{center}
    \begin{figure}[htbp]
    		\includegraphics[clip, width=\linewidth]{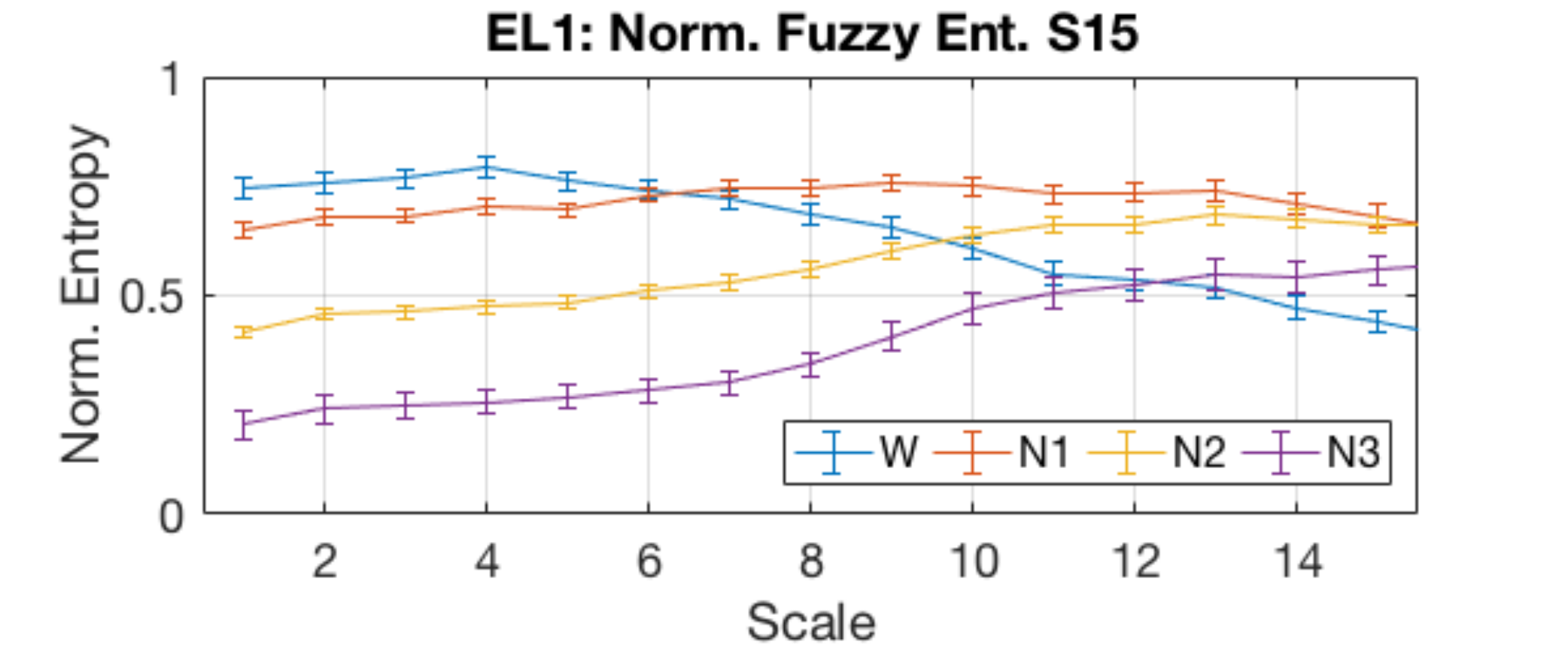}
    	\caption{Structural complexity features for different sleep stages. Normalised multi-scale fuzzy entropy (MSFE) from the in-ear EEG channel EL1 is evaluated the over scales 1 (normal FE) to 15, and shows excellent separation between sleep stages. The error bars indicate the standard error.}
	\label{fig:N_MSFE}
	\end{figure}
\end{center}

The multi-scale entropy (MSE) method calculates structural complexity of time-series over multiple temporal scales \cite{Costa2002, Ahmed2011}, and can be measured with e.g. sample entropy, approximate entropy, and permutation entropy. We used multi-scale fuzzy entropy (MSFE) \cite{Ahmed2016} with a small embedding dimension, owing to its robustness in the presence of noise. The following parameters for MSFE were chosen: maximum scale $\tau = 15$, $m = 2$, $n = 2$, $r = 0.15 \times$({\it standard deviation of each epoch}). Overall, 15 features were extracted from the EL1 channel and were normalised, as illustrated in Figure \ref{fig:N_MSFE}. Observe the good separation of entropy values between sleep stages in each scale; in particular, structural complexity for the Wake condition decreased with the scale factor. For the N3 sleep stage, a large proportion of power is contained in the delta band (relative to total power), and this more deterministic behaviour caused  the FE values to be smaller than in other sleep stages.

\subsection{Classification} 
The classification was performed based on 30 SEF and MSFE features, which were normalised to the range $[0 \,\, 1]$. The one-against-one multi-class support vector machine (SVM) with a radial basis function (RBF) kernel was employed as a classifier \cite{Chang2011}.

\section{Results}
\subsection{Performance Evaluation}
Feature extraction was performed using Matlab 2016b, and the classification was conducted in Python 2.7.12 Anaconda 4.2.0 (x86\_64) operated on an iMac with 2.8GHz Intel Core i5, 16GB of RAM. A 5-fold cross validation (CV) was performed to evaluate the automatic sleep stage classification. The performance metrics used were class-specific sensitivity (SE) and precision (PR), as well as overall accuracy (AC) and Kappa coefficients ($\kappa$), defined as follows:
\begin{eqnarray*}
SE = \frac{TP}{TP+FN}, \,\, PR = \frac{TP}{TP+FP}, \,\, AC = \frac{\sum_{i = 1}^{C} TP_i}{N}, \\
\pi_e = \frac{\sum_{i = 1}^{C} \left\{(TP_i + FP_i) (TP_i + FN_i)\right\}}{N^2}, \,\, \kappa = \frac{AC - \pi_e}{1 - \pi_e}.
\end{eqnarray*}
The parameter TP (true positive) represents the number of positive (target) epochs correctly predicted, TN (true negative) is the number of negative (non-target) epochs correctly predicted, FP (false positive) is the number of negative epochs incorrectly predicted as positive class, FN (false negative) is the number of positive epochs incorrectly predicted as negative class, $C$ is the number of classes, and $N$ the total number of epochs.

\subsection{Scenario1: Sleep stage classification from ear-EEG against the manually scored hypnogram based on ear-EEG}
We first evaluated the agreement between the hypnogram scored based on ear-EEG channels and the predicted label based on extracted features from the in-ear EEG channel EL1. Tables \ref{table:2class_score_ear}, \ref{table:2class_deep_score_ear}, \ref{table:4class_score_ear} show the confusion matrices obtained from the classification results based on the SEF and MSFE features for the 2-class scenarios Wake vs Sleep and W-N1 vs N2-N3, and the 4-class (W, N1, N2, N3) scenario. For the 2-class classification scenarios, the overall classification accuracies were \unit[95.2]{\%} and \unit[86.0]{\%} with an Almost Perfect ($\kappa = 0.83$) to Substantial ($\kappa = 0.68$) agreement of Cohen's Kappa coefficients \cite{Landis2008}, as shown in Table \ref{table:2class_score_ear} and \ref{table:2class_deep_score_ear}.

\begin{table}[htbp]
\centering
\caption{Confusion matrix for the 2-class Wake vs Sleep classification}
\label{table:2class_score_ear}
\begin{tabular}{P{1.3cm} P{0.5cm} P{1.5cm} P{1.5cm}  P{1.3cm}}
\hline
  \multicolumn{2}{c}{\multirow{2}{*}{} }                          & \multicolumn{2}{c}{Algorithm based on ear-EEG}    \\
               &             & Wake & Sleep & SE / PR\\
 \hline
 \multirow{2}{*}{\parbox{1.5cm}{Score based on ear-EEG}} & Wake &42 & 10   & 80.8 / 91.3\\
 								& Sleep     &4  &237  & 98.3 / 96.0  \\
\hline
\multicolumn{5}{c}{Accuracy: \unit[95.2]{\%}, Kappa = 0.83} \\
              \hline
              \end{tabular}
\end{table}
\begin{table}[htbp]
\centering
\caption{Confusion matrix for the 2-class Wake-N1 vs N2-N3 classification}
\label{table:2class_deep_score_ear}
\begin{tabular}{P{1.3cm} P{0.8cm} P{1.5cm} P{1.5cm}  P{1.3cm}}
\hline
  \multicolumn{2}{c}{\multirow{2}{*}{} }                          & \multicolumn{2}{c}{Algorithm based on ear-EEG}    \\
               &             & W-N1 & N2-N3 & SE / PR\\
 \hline
 \multirow{2}{*}{\parbox{1.5cm}{Score based on ear-EEG}}
 & W-N1 &76 & 25   & 75.3 / 82.6\\
 & N2-N3     &16  &176  & 91.7 / 87.6  \\
\hline
\multicolumn{5}{c}{Accuracy: \unit[86.0]{\%}, Kappa = 0.68} \\
              \hline
              \end{tabular}
\end{table}

The accuracy for the more difficult 4-class sleep stage classification was \unit[78.5]{\%} with the Kappa coefficient $\kappa = 0.64$, which indicates a Substantial agreement, as shown in Table \ref{table:4class_score_ear}. 

\begin{table}[htbp]
\centering
\caption{Confusion matrix for 4-class sleep stage classification}
\label{table:4class_score_ear}
\begin{tabular}{P{1.3cm} P{0.5cm} P{0.5cm} P{0.5cm} P{0.5cm} P{0.5cm}  P{1.3cm}}
\hline
  \multicolumn{2}{c}{\multirow{2}{*}{} }                          & \multicolumn{4}{c}{Algorithm based on ear-EEG}    \\
               &             & W & N1 & N2 & N3 & SE / PR\\
 \hline
 \multirow{4}{*}{\parbox{1.5cm}{Score based on ear-EEG}}
 & W      &46 & 3  &3 & 0 & 88.5 / 92.0\\
 & N1     &3  &17 & 27 & 2& 34.7 / 56.7  \\
  & N2     &1  &10 &146 &5  & 90.1 / 78.9  \\
 & N3    &0  &0 & 9& 21  & 70.0 / 75.0  \\
\hline
\multicolumn{7}{c}{Accuracy: \unit[78.5]{\%}, Kappa = 0.64} \\
              \hline
              \end{tabular}
\end{table}

\subsection{Scenario2: Sleep stage classification from ear-EEG against the manually scored hypnogram based on scalp-EEG}
We next evaluated the agreement between the hypnogram scored based on scalp-EEG channels and the predicted label based on extracted features from the in-ear EEG channel EL1. Tables \ref{table:2class_score}, \ref{table:2class_deep_score}, \ref{table:4class_score} show the corresponding confusion matrices obtained from classification results with the SEF and MSFE features for the 2-class Wake vs Sleep and W-N1 vs N2-N3 scenarios, and the 4-class (W, N1, N2, N3) scenario. For the 2-class classification problems, the achieved classification accuracies were more than \unit[90]{\%}, with the Substantial ($\kappa = 0.75$) to Almost Perfect ($\kappa = 0.80$) values of the Kappa coefficients \cite{Landis2008}. 
\begin{table}[htbp]
\centering
\caption{Confusion matrix for the 2-class Wake vs Sleep classification}
\label{table:2class_score}
\begin{tabular}{P{1.8cm} P{0.5cm} P{1.3cm} P{1.3cm}  P{1.3cm}}
\hline
  \multicolumn{2}{c}{\multirow{2}{*}{} }                          & \multicolumn{2}{c}{Algorithm based on ear-EEG}    \\
               &             & Wake & Sleep & SE / PR\\
 \hline
 \multirow{2}{*}{\parbox{1.8cm}{Score based on scalp-EEG}} 
 & Wake &50 & 17   & 74.6 / 87.7\\
& Sleep     &7  &219  & 96.9 / 92.8  \\
\hline
\multicolumn{5}{c}{Accuracy: \unit[91.8]{\%}, Kappa = 0.75} \\
              \hline
              \end{tabular}
\end{table}
\begin{table}[htbp]
\centering
\caption{Confusion matrix for the 2-class Wake-N1 vs N2-N3 classification}
\label{table:2class_deep_score}
\begin{tabular}{P{1.7cm} P{0.8cm} P{1.1cm} P{1.1cm}  P{1.3cm}}
\hline
  \multicolumn{2}{c}{\multirow{2}{*}{} }                          & \multicolumn{2}{c}{Algorithm based on ear-EEG}    \\
               &             & W-N1 & N2-N3 & SE / PR\\
 \hline
 \multirow{2}{*}{\parbox{1.7cm}{Score based on scalp-EEG}}
 & W-N1 &96 & 17   & 85.0 / 90.0\\
& N2-N3     &11  &169  & 89.7 / 90.9  \\
\hline
\multicolumn{5}{c}{Accuracy: \unit[90.4]{\%}, Kappa = 0.80} \\
              \hline
              \end{tabular}
\end{table}

The achieved accuracy for the 4-class sleep stage classification was \unit[76.8]{\%}, with the kappa coefficient $\kappa = 0.65$, which indicates a Substantial agreement. 
\begin{table}[htbp]
\centering
\caption{Confusion matrix for the 4-class sleep stage classification}
\label{table:4class_score}
\begin{tabular}{P{1.8cm} P{0.5cm} P{0.5cm} P{0.5cm} P{0.5cm} P{0.5cm}  P{1.3cm}}
\hline
  \multicolumn{2}{c}{\multirow{2}{*}{} }                          & \multicolumn{4}{c}{Algorithm based on ear-EEG}    \\
               &             & W & N1 & N2 & N3 & SE / PR\\
 \hline
 \multirow{4}{*}{\parbox{1.8cm}{Score based on scalp-EEG}}
& W      &53 & 5  &8 & 1 & 79.1 / 77.9\\
& N1     &12  &23 & 11 & 0& 50.0 / 65.7  \\
& N2     &3  &5 &125 &7  & 89.3 / 79.1  \\
& N3    &0  &2 & 14& 24  & 60.0 / 75.0  \\
\hline
\multicolumn{7}{c}{Accuracy: \unit[76.8]{\%}, Kappa = 0.65} \\
              \hline
              \end{tabular}
\end{table}
\begin{center}
    \begin{figure}[h]
		\includegraphics[clip, width=\linewidth]{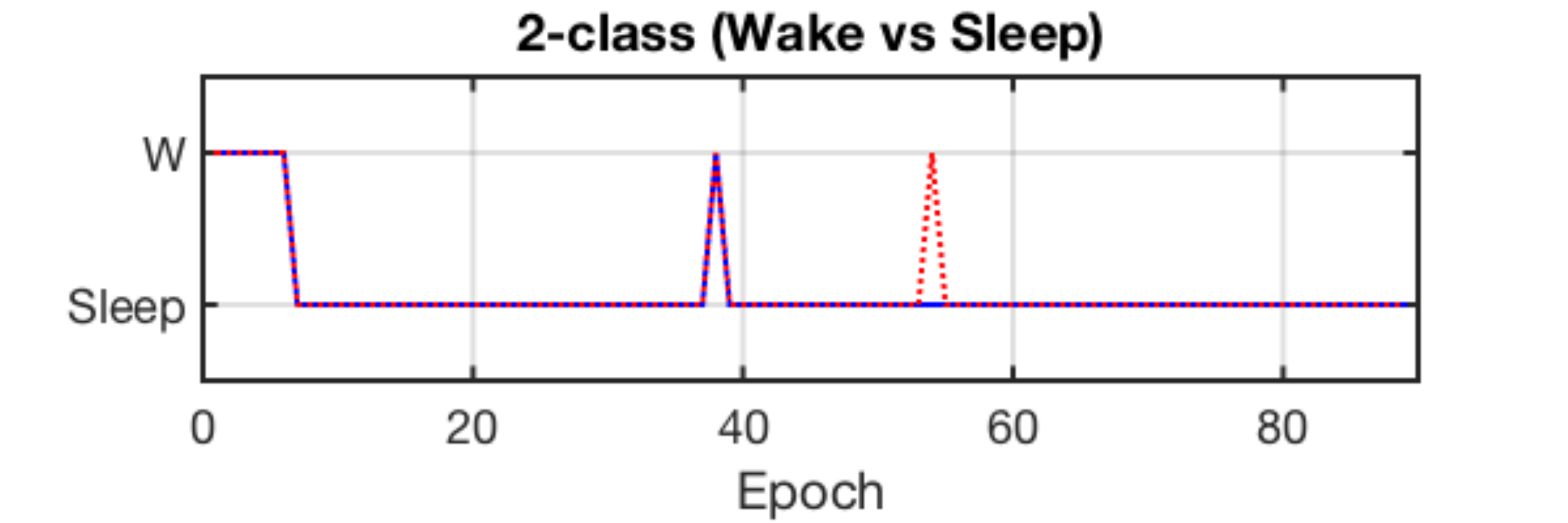}
		\includegraphics[clip, width=\linewidth]{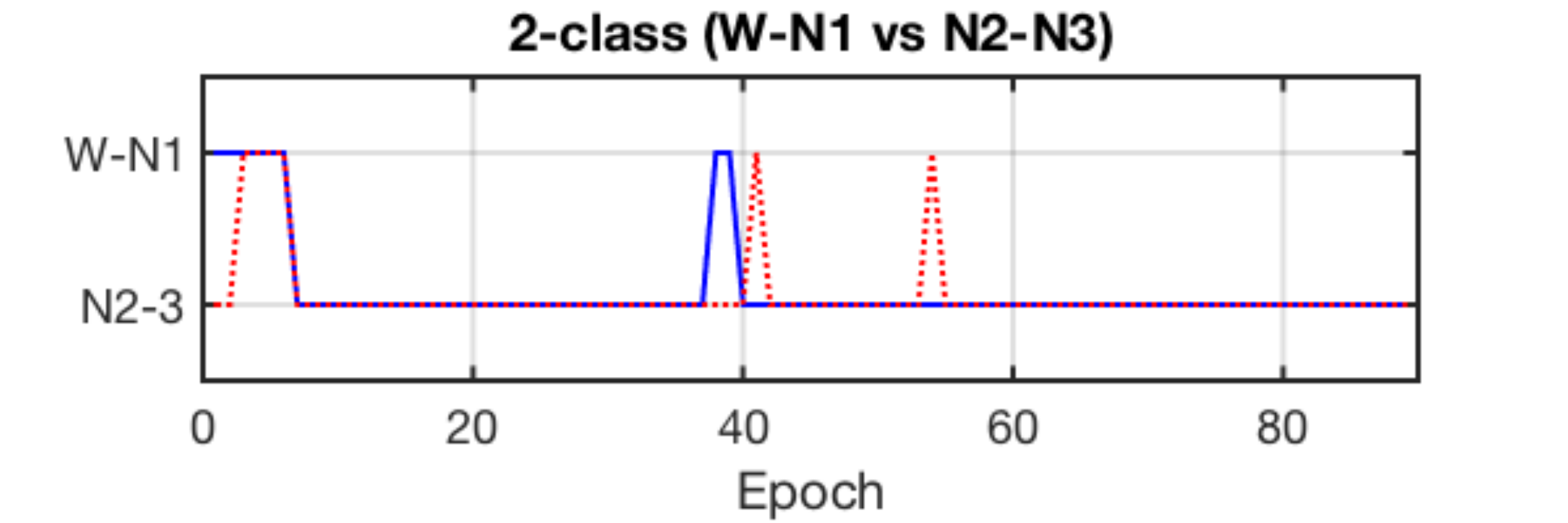}
		\includegraphics[clip, width=\linewidth]{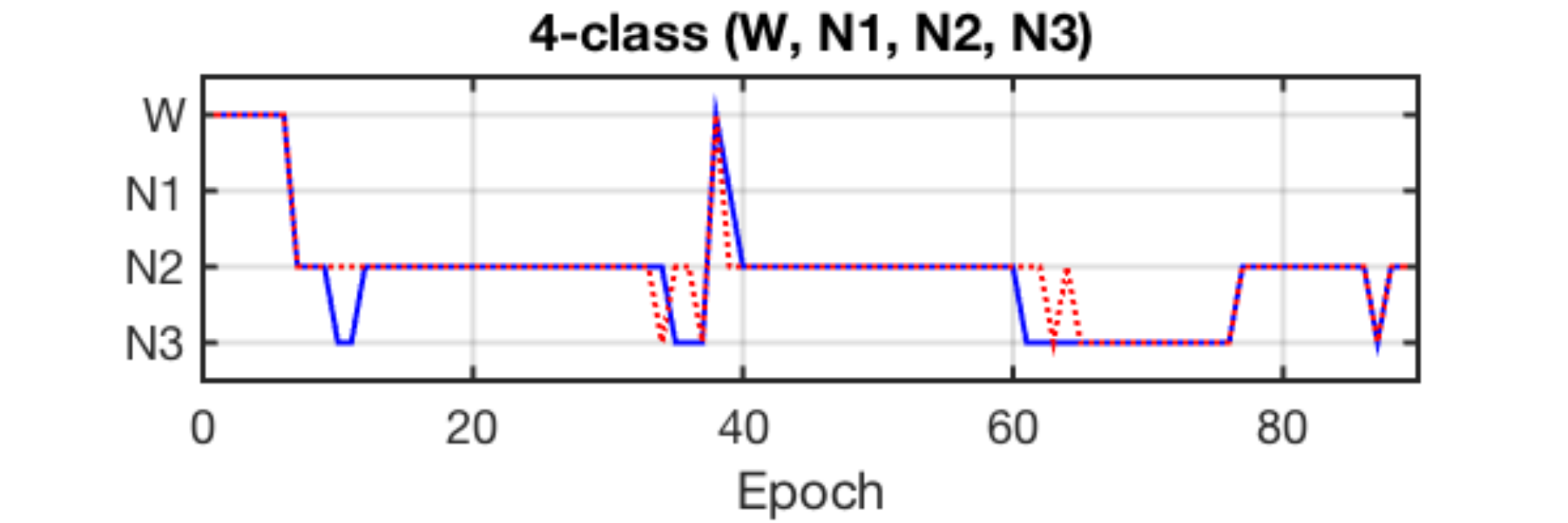}

    	\caption{Hypnogram for Subject 2 scored based on scalp-EEG channels (blue) and the automatically predicted label based on in-ear EEG channel EL1 (red) for the 2-class Wake vs Sleep (top) and W-N1 vs N2-N3 (middle) scenarios, and the 4-class (bottom) classification scenario.}
	\label{fig:HypResults}
    \end{figure}
\end{center}

Figure \ref{fig:HypResults} depicts the hypnograms scored manually based on scalp-EEG channels (blue) and the automatically predicted label based on the in-ear EL1 channel (red) for the 2-class Wake vs Sleep (top) and W-N1 vs N2-N3 (middle) scenarios, and the 4-class (bottom) scenario, for the Subject 2. Only the first epoch was removed because of the AC onset noise, therefore the hypnogram was scored based on 89 epochs, which corresponds to 44 minutes of \unit[30]{s} recording. 
For the 4-class problems, even though some epochs were predicted incorrectly, for example epoch 62 (hypnogram:N3, prediction:N2), the majority of epochs were correctly classified.  
This confirms that the features extracted from the ear-EEG data were effectively used for the automatic sleep stage classification, and provided a substantial match to the scalp-EEG patterns scored manually by an expert. We can therefore conclude that the recorded ear-EEG carried a sufficient amount of information to evaluate human sleep robustly.

\subsection{Agreement between the predicted and manual sleep scores}
Upon establishing the feasibility of predicting scalp-EEG sleep stages from ear-EEG features, we shall now benchmark these findings against our recent results based on manual scoring of both scalp- and ear-EEG \cite{Looney2016a}. To this end, Table \ref{table:comparison} compares the manual and automatic labels for the following scenarios:
\begin{itemize}
 \item \emph{Scenario 1}: The manually scored hypnogram based on ear-EEG channels vs the predicted label based on the in-ear EL1 channel (Table \ref{table:2class_score_ear}, \ref{table:2class_deep_score_ear}, \ref{table:4class_score_ear}).
 \item \emph{Scenario 2}: The manually scored hypnogram based on scalp-EEG channels vs the predicted label based on the in-ear EL1 channel (Table \ref{table:2class_score}, \ref{table:2class_deep_score}, \ref{table:4class_score}).
 \item The hypnogram manually scored based on scalp-EEG channels vs that scored based on ear-EEG channels. 
 \end{itemize}
In all cases, the proposed automatically scored labels were a significant match to the corresponding labels scored manually in \cite{Looney2016a}. 

\begin{table}[htbp]
\centering
\caption{Comparison between the manual scores and automatic predicted scores (Accuracy [\%] / Kappa)}
\label{table:comparison}
\begin{tabular}{cccc}
\hline
               &   \emph{Scenario 1}         & \emph{Scenario 2}         & \cite{Looney2016a}\\
               \hline
               &   Ear label         & Scalp label          & Scalp label \\
               &  vs                       & vs                     &  vs \\
               &  Ear Prediction     & Ear Prediction &  Ear label  \\
 
 \hline
Wake vs Sleep  & 95.2 / 0.83 & 91.8 / 0.75 & 84.0 / 0.60\\
W-N1 vs N2-N3 & 86.0 / 0.68 & 90.4 / 0.80  & 83.0 / 0.65\\
4 class               & 78.5 / 0.64 & 76.8 / 0.65  & - / -\\
\hline
\end{tabular} 
\end{table}


\section{Conclusions}
We have established the feasibility of automatic sleep scoring based on an in-ear EEG sensor. For rigour, this has been achieved for two scenarios: \emph{Scenario 1} examined automatic scores for ear-EEG against manual scores for ear-EEG, while \emph{Scenario 2} examined automatic scores for ear-EEG against manual scores for scalp-EEG. The so performed sleep stage prediction from ear-EEG for the 2-class sleep stage classification (Wake vs Sleep and W-N1 vs N2-N3) for \emph{Scenario 1} gave the overall accuracy of \unit[95.2]{\%} and \unit[86.0]{\%} with the corresponding Kappa coefficients of 0.83 and 0.68, which indicates Almost Perfect and Substantial agreements. For the 4-stage classification the accuracy was \unit[78.5]{\%} with $\kappa=0.64$, a Substantial agreement.
For \emph{Scenario 2}, the corresponding accuracies for the 2-stage classification were \unit[91.8]{\%} and \unit[90.4]{\%} with the Kappa coefficient $\kappa=0.75$ and $\kappa=0.80$ (Substantial to Almost Perfect agreement), while for the 4-stage classification the accuracy was \unit[76.8]{\%} with $\kappa=0.65$, a Substantial agreement.  We have therefore confirmed both empirically and over comprehensive statistical testing that the in-ear EEG carries sufficient amount of information to faithfully represent human sleep patterns, thus opening up a new avenue in fully wearable sleep research. For this pilot study the number of subjects was only four, and our future studies will consider a larger cohort of subjects, overnight sleep, and other aspects of fully wearable scenarios. 

\bibliographystyle{ieeetr}
\bibliography{library_comp}


\end{document}